\documentclass[conference]{IEEEtran}
\usepackage[pdftex]{graphicx}

\usepackage[caption=false,font=footnotesize]{subfig}

\usepackage{amsmath}

\usepackage{cite}

\usepackage{url}

\usepackage{siunitx}

\newcommand{\rrx}{r^{\text{Rx}}}
\newcommand{\rtx}{r^{\text{Tx}}}

\newcommand{\prt}{p(r,t|r_0)}
\newcommand{\prtzero}{p(r,t \rightarrow 0 |r_0)}

\newcommand{\tend}{t_{\text{end}}}

\DeclareMathOperator*{\argminn}{\arg\min}

\begin{document}

 
 \title{A Machine Learning Approach to Model the Received Signal in Molecular Communications}

\author{\IEEEauthorblockN{H. Birkan Yilmaz, Changmin Lee, Yae Jee Cho, and Chan-Byoung Chae}
\IEEEauthorblockA{School of Integrated Technology,  Institute of Convergence Technology\\
Yonsei University, Korea. \\ 
Email:\{birkan.yilmaz, cm.lee, yjenncho, cbchae\}@yonsei.ac.kr}%
}


\maketitle

\begin{abstract}
A molecular communication channel is determined by the received signal. Received signal models form the basis for studies focused on modulation, receiver design, capacity, and coding depend on the received signal models. Therefore, it is crucial to model the number of received molecules until time $t$ analytically. Modeling the diffusion-based molecular communication channel with the first-hitting process is an open issue for a spherical transmitter. In this paper, we utilize the artificial neural networks technique to model the received signal for a spherical transmitter and a perfectly absorbing receiver (i.e., first hitting process). The proposed technique may be utilized in other studies that assume a spherical transmitter instead of a point transmitter. 
\end{abstract}


\IEEEpeerreviewmaketitle

\section{Introduction}
The manipulation of matter at the atomic and the molecular scale constitutes nanotechnology. It is a promising technology that has numerous potential applications~\cite{farsad2016comprehensiveSO,nakano2013molecularC_BOOK,akyildiz2011nanonetworksAN}. One of its  innovative approaches is to utilize collaborative behavior amongst small entities. To enable the revolutionary possibilities of nanotechnology, it is important to possess the capacity to communication at the nano- and micro-scale~\cite{guo2015molecularCC_WCMAG}. As a possible means to communication at such a small scale, researchers have proposed molecular communication via diffusion (MCvD)~\cite{hiyama2005molecularC}. Examples of MCvD may be found prevalently in nature--quorum sensing between bacteria at micro-scale and pheromone-based communication between animals at macro-scale in both air and water environments. What occurs in these examples is molecules are utilized to convey information~\cite{farsad2016comprehensiveSO,nakano2013molecularC_BOOK,akyildiz2011nanonetworksAN,gine2009molecularCO,gregory2010newNA_bacteria_JSAC,agosta1992chemicalCT,berg1993random,lio2012opportunisticRT,kuran2010energyMF}. In an MCvD system, a transmitter node emits molecules and molecules propagate through the environment until they arrive at the receiver node. Received molecules constitute the received signal. This is of prime importance when it comes to modeling an MCvD channel.

One of the main challenges in MC is to develop valid models for representing the received signal in different environments and conditions. Some of the MCvD models in the literature, assume that whenever a molecule hits the receiver it is removed from the environment~\cite{kuran2010energyMF,srinivas2012molecularCI_inverseG,kadlor2012molecularCU_drift_TNBS,nakano2012channelMA_oneD_COML,yilmaz2014threeDC,yilmaz2014simulationSO_SIMPAT}. This phenomena is modeled by the first-passage process (a.k.a first-hitting process). In this model, each molecule can contribute to the received signal only once. In~\cite{srinivas2012molecularCI_inverseG}, the authors presented the analytical model for the received signal in a 1-dimensional (1D) environment while considering the first-passage process. In~\cite{kadlor2012molecularCU_drift_TNBS}, the authors enhanced the model from~\cite{srinivas2012molecularCI_inverseG} by incorporating the drift component.  In~\cite{yilmaz2014threeDC}, the authors presented the expected cumulative number of received molecules when the transmitter is a point source and the receiver is an absorbing spherical node in a 3D medium. In~\cite{yilmaz2014arrivalMF_EL}, the authors compared the arrival process models used with the expected received signal that was derived in~\cite{yilmaz2014threeDC}. In \cite{akkaya2015effectOR_receptor_COML}, the authors analytically modeled the received signal and derived the expected cumulative number of received molecules when the receptor effect was added to the system presented in~\cite{yilmaz2014threeDC}. 

On the other hand, some of the MCvD models ignore the first-passage process, allowing molecules to pass through the receiver node surface/boundary with no interaction between the environment and the receiver~\cite{noel2014improvingRP_enzymes_TNBS,kilinc2013receiverDF_JSAC,pierobon2010physicalET_JSAC}. In such models, the molecules are allowed to contribute to the signal multiple times, as they can pass in and out of the receiver node surface in and out multiple times. 

Both of the MCvD physical layer models are summarized and presented in~\cite{guo2015molecularCC_WCMAG}. Researchers~\cite{guo2015molecularCC_WCMAG,cuatrecasas1974membraneR} have claimed that the first-hitting process is observed in nature more frequently than the passive receiver process~\cite{guo2015molecularCC_WCMAG,cuatrecasas1974membraneR}. Moreover, some of these models are validated by macro-scale testbed implementations~\cite{farsad2014ChannelAN_noise_JSAC,lee2015molecularMIMO_INFOCOM,koo2016molecularMIMO_JSAC}. Both channel models (i.e., the received signal for the first-hitting process and passive receiver) have a common hurdle for the communication engineering: heavy tail distribution of the received signal, which causes inter-symbol-interference (ISI). In literature, plenty of techniques are proposed to eliminate the severe effects of ISI by utilizing specialized modulations~\cite{kuran2011modulationTF_ICC,kuran2012interferenceEO,tepekule2014energyEI_MTSK,tepekule2015isiMT_TMBMC,kim2013novelMT_isomers_JSAC,kim2014symbolIO_TNBS,guo2015molecularCW_passband}, error correcting codes~\cite{leeson2012forwardEC_hamming,mahfuz2013performanceAO_convolutional,shih2013channelCF_zeroErr}, and  enzymes~\cite{kuran2013tunnelBA,noel2014improvingRP_enzymes_TNBS,heren2015effectOD_degradation_TMBMC,yilmaz2016interferenceRV_enzymeDeployment_EL,cho2016effectiveIS_enzymeDeployment_ETT}. In these studies, analytical models and their assumptions are used for the derivations, which indicates the importance of modeling the channel and the  received signal in different environments. 

Modeling the MCvD channel with the first-hitting process is an open issue for a spherical transmitter. The literature introduces analytical models for the first-hitting process with a point transmitter and with a spherical receiver~\cite{yilmaz2014threeDC}. In this paper, we propose a channel model that is inspired from the derived formulation in~\cite{yilmaz2014threeDC}. The proposed model function has model parameters that are affected by the system parameters. Moreover, we show that these model parameters can be learned by an artificial neural network (ANN) and the trained network can predict the model parameters for a given system parameters.

\section{System Model}
\begin{figure}[!t]
\begin{center}
	\includegraphics[width=1.0\columnwidth,keepaspectratio]%
	{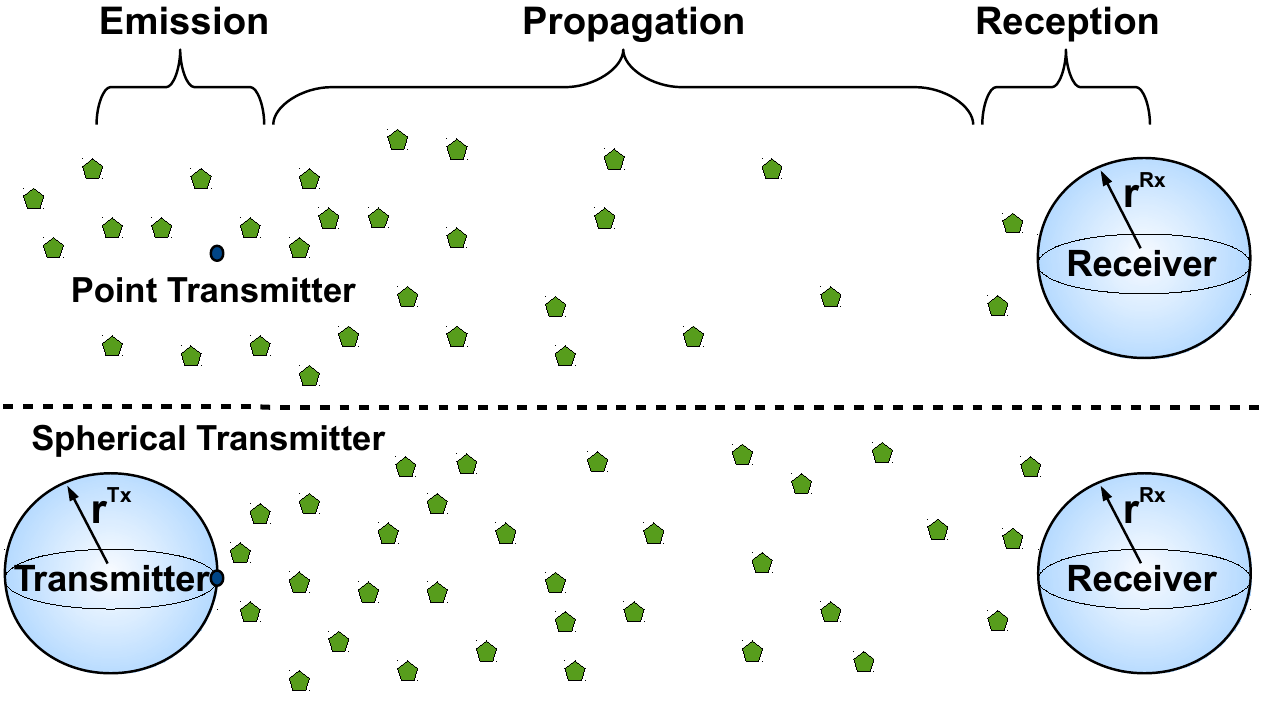}
	\caption{System model of MCvD with point and spherical transmitter cases. In the point transmitter case, molecules are free to go in the opposite direction. On the other hand, molecules are reflected by the body of the transmitter in the spherical transmitter case.}
	\label{fig_system_modell}
\end{center}
\end{figure}
In an MCvD system, there is at least one transmitter and receiver pair in a fluid environment. Figure~\ref{fig_system_modell} shows two different cases for MCvD--point and spherical transmitters. In general, the point transmitter case is studied in the literature~\cite{yilmaz2014threeDC,pierobon2010physicalET_JSAC,akkaya2015effectOR_receptor_COML}. The point transmitter assumption are reasonable for some applications. However the transmitter node has a body in general and the transmitter does not react to the molecules from itself. In this paper, we model the MCvD channel with a perfectly absorbing receiver (i.e., whenever a molecule hits the surface of the receiver node, it contributes to the received signal) without the point transmitter assumption.   

As shown in Fig.~\ref{fig_system_modell}, emitted molecules diffuse in the 3D environment, which is characterized by the diffusion coefficient~$D$. At the receiver side, the radius of the receiver is denoted by $\rrx$ and the received signal consists of the time histogram of hitting molecules. When we have a point transmitter, molecules are able to more freely travel in the opposite direction of the receiver more freely. In the spherical transmitter case, however, molecules are obstructed and reflected by the transmitter of radius $\rtx$. Hence, the received signals of these two cases are expected to differ.

\begin{figure*}[!t]
\begin{center}
	\includegraphics[width=1.99\columnwidth,keepaspectratio]%
	{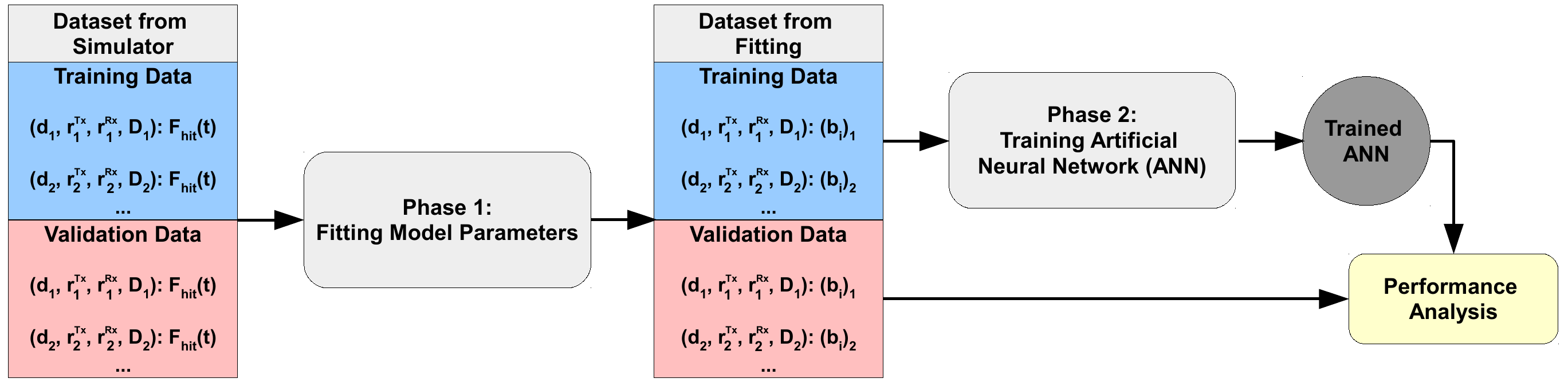}
	\caption{Flowchart of the proposed technique. Phase 1 deals with fitting model parameters by utilizing the dataset from simulator. Phase 2 deals with training an ANN on the training dataset that is obtained from Phase 1. Dataset from Phase 1 consists of input-output pairs where input is ${(d,\, \rtx,\, \rrx,\, D)}$ and the output is the model parameters (i.e., $b_i$'s).}
	\label{fig_proposed_technq_flowchart}
\end{center}
\end{figure*}

\subsection{MCvD with a Point Transmitter}
The received signal is modeled analytically for the point transmitter case. The diffusion process basically models the average movement of particles in the concentration gradient. For modeling the received signal, we need to include the boundary conditions that represent the concentration and molecule distribution function. First, we consider Fick's Second Law in a 3D environment, which is given by
\begin{align}
\label{eqn_ficks3d}
\frac{\partial \prt}{\partial t} = D \nabla^2 \prt
\end{align}
where $\nabla^2$, $\prt$, and $D$ are the Laplacian operator, the molecule distribution function at time $t$ and distance $r$ given the initial distance $r_0$, and the diffusion constant. The initial condition is given by
\begin{align}
\label{eqn_initial_condition}
\prtzero = \frac{1}{4 \pi r_0^2} \delta(r - r_0),
\end{align}
and the boundary conditions by
\begin{align}
\label{eqn_bd_condition_one}
\lim_{r \rightarrow \infty} \prt = 0,
\end{align}
\begin{align}
\label{eqn_bd_condition_two}
D \frac{\partial \prt}{\partial r} = w \,\prt \text{, for } r = \rrx
\end{align}
where $\rrx$ and $w$ denote the radius of the receiver and the rate of reaction. The reaction rate with the receiver boundary is controlled by $w$.  Specifically, $w\!=\!0$ means a  nonreactive surface and $w\!\rightarrow\!\infty$, on the other hand, corresponds to the boundary where every collision leads to an absorption. 

In~\cite{yilmaz2014threeDC}, the solution to this differential equation system is presented and analyzed from the perspective of channel characteristics. After finding the reaction rate, the authors presented the formula for the fraction of molecules that hit the receiver until time $t$, as follows:
\begin{align}
\begin{split}
F_\text{hit}^{3\text{D}} (t)=  \frac{\rrx}{d\!+\!\rrx} \,\text{erfc} \left( \frac{d}{\sqrt{4Dt\,}}\right) = \frac{2 \rrx}{d\!+\!\rrx} \,\Phi\left(\frac{-d}{\sqrt{2Dt}}\right)
\end{split}
\label{eqn_3d_frac_received_point_src}
\end{align}
where $d$, $\text{erfc}(.)$, and $\Phi(.)$ represent the distance, complementary error function, and the standard Gaussian cdf, respectively. When the transmitter is a point, there is a circular symmetry, all the points at the same radius are equivalent and the solution for the system of differential equations is enabled. For the spherical transmitter case, however, it is more complex to derive the formulation of the number of received molecules.

\subsection{MCvD with a Spherical Transmitter}
Modeling the received signal analytically for a spherical transmitter is an open issue when the receiver is a perfectly absorbing spherical receiver in a 3D environment. The main difference and  hurdle stem from the lack of circular symmetry. For the spherical transmitter case, molecules are biased towards going in the direction of the receiver due to the obstructing body of the transmitter node. Therefore, each of the molecules is expected to have a higher probability of hitting the receiver. 

Hurdles caused by a spherical transmitter steered us to simulate the MCvD with a spherical transmitter and to analyze the patterns so as to grasp the underlying dynamics. We ran extensive simulations with many parameters. We realized that the cumulative number of received molecules exhibits a similar behavior to the point transmitter case with a small perturbation that is dependent on the system parameters.

\section{Proposed Technique for Channel Modeling}
As noted above, the main challenge in MCvD is to develop valid models for representing the received signal in different environments and conditions. When the transmitter has a spherical body, we propose to model the received signal by parameterizing \eqref{eqn_3d_frac_received_point_src} and learning the patterns behind the model parameters.

The proposed technique is composed of two main phases: fitting the model parameters constitutes phase one (i.e., forming the input-output dataset for phase two) and learning the patterns in the input-output dataset constitutes phase two. After the learning phase (i.e., phase two), the output of the algorithm is a trained ANN for future predictions on unexplored input. A representative scheme of the proposed technique is depicted in Fig.~\ref{fig_proposed_technq_flowchart}.

\subsection{Model Function and Fitting}
We propose two different model functions for fitting the simulation data and name them as \emph{primitive model} and \emph{enhanced model}. We use only a scaling factor for the primitive model, which is represented as follows:
\begin{align}
\begin{split}
F_\text{hit}^{3\text{D}} (t, b_1)=  b_1 \, \frac{\rrx}{d\!+\!\rrx} \,\text{erfc} \left( \frac{d}{\sqrt{4Dt\,}}\right) 
\end{split}
\label{eqn_model_primitive}
\end{align}
where $b_1$ represents the model fitting parameter. For the enhanced model, we also parametrize the components related to $D$ and $t$, which is shown as follows:
\begin{align}
\begin{split}
F_\text{hit}^{3\text{D}} (t, b_1, b_2, b_3)=  b_1 \, \frac{\rrx}{d\!+\!\rrx} \,\text{erfc} \left( \frac{d}{(4D)^{b_2} \, t^{b_3}}\right) 
\end{split}
\label{eqn_model_enhanced}
\end{align}
where $b_1$, $b_2$, and $b_3$ are model fitting parameters. These model fitting parameters are introduced for fitting simulation data to model functions.

To find the model parameters corresponding to the system parameters, we use nonlinear least squares curve fitting technique. Assuming that we have $N$ observations during the simulation we formulate the  fitting model parameter estimation problem with $m$ parameters as follows:
\begin{align}
\begin{split}
\argminn\limits_{b_1,...,b_m} \sum\limits_{k=1}^{N} \left(F_\text{hit}^{3\text{D}} (t_k, b_1,...,b_m)- S_\text{hit}^{3\text{D}} (t_k) \right)^2
\end{split}
\label{eqn_fitting_lsq}
\end{align}
where $S_\text{hit}^{3\text{D}} (t)$ corresponds to simulation data that is representing the ratio of hitting molecules until time $t$.

For primitive and enhanced models, we apply to the simulation data nonlinear least squares curve-fitting technique. The output of the curve fitting process consists of the model parameters (i.e., $b_1$ for primitive model and $b_1 \!\sim\! b_3$ for the enhanced model). Hence, we obtain model parameters for each simulation case, which forms the dataset of the next phase. This dataset structure is depicted in Fig.~\ref{fig_proposed_technq_flowchart} in the middle panel, which contains simulation case system parameters and the model parameters (i.e., $b_i$'s) from curve fitting. 

\begin{table}[!t]
\begin{center}
\caption{Range of parameters used in the experiments}
\renewcommand{\arraystretch}{1.14}
\label{tbl_system_parameters}
\begin{tabular}{p{5.2cm} l}
\hline
\bfseries{Parameter} 							& \bfseries{Value} \\ 
\hline 
Number of emitted molecules			& $3\,000$\\
Simulation duration ($\tend$) 		& $\SI{1}{\second}$\\ 
Replication 						& $500$\\
TDS Distances ($d$) 				& $\{2,\, 4,\, 6,\, 8,\, 10\}\,\, \si{\micro\metre} $\\
VDS Distances ($d$) 				& $\{3,\, 5,\, 7,\, 9,\, 11\}\,\, \si{\micro\metre} $\\
TDS Transmitter radii ($\rtx$)		& $\{5,\, 7.5,\, 10\}\,\,\si{\micro\metre}$\\
VDS Transmitter radii ($\rtx$)		& $\{4,\, 6,\, 8\}\,\,\si{\micro\metre}$\\
TDS Diffusion coefficients ($D$) 	& $\{50, \, 75, \, 100\}\si{\micro\metre^2/\second}$\\
VDS Diffusion coefficients ($D$) 	& $\{60, \, 70, \, 80\}\si{\micro\metre^2/\second}$\\
TDS Receiver radii ($\rrx$)			& $\{5,\, 7.5,\, 10\}\,\,\si{\micro\metre}$\\
VDS Receiver radii ($\rrx$)			& $\{4,\, 6,\, 8\}\,\,\si{\micro\metre}$\\
\hline
\end{tabular} 
\end{center}
\renewcommand{\arraystretch}{1}
\end{table}

\subsection{Learning Model Parameters}
In this paper, we introduce a machine-learning technique to model the received signal in MCvD with a spherical transmitter. One of the popular machine-learning techniques is artificial neural networks. They have simple neuron-like nodes with thresholds and the connections with weights. Basically, the thresholds and the weights are adjusted until the desired output is observed for the given inputs. 

The dataset from the curve fitting phase is divided into two disjointed subsets as training and validation  datasets. Training data is utilized for training the ANN for getting the desired output for given inputs. Bayesian regularization backpropagation technique is used for the ANN training that updates the weights and bias values according to Levenberg-Marquardt optimization. Bayesian regularization minimizes a combination of squared errors and weights to determine the ANN parameters that generalize the pattern in the input-output pairs. We utilize the trained ANN to estimate the channel parameters for different cases. Note that the curve-fitting technique requires simulation data but a trained ANN does not require any simulation data, i.e., required inputs are the system parameters such as $d$, $\rtx$, $D$, and $\rrx$.

\section{Results and Analysis}

\subsection{Performance Metrics and Parameters}
For the performance metrics, we mainly use root mean squared error (RMSE) with respect to simulation data in terms of number of received molecules. First, we give the received signal and signal-to-interference (SIR) plots for the example cases. We then present the average RMSE over different cases. 

Common system parameters for simulation, training (TDS), and validation (VDS) datasets are presented in Table~\ref{tbl_system_parameters}. From the given datasets, each of the VDS and TDS have ${5\times 3\times 3\times 3 \!=\!135}$ different cases, making a total of $270$ cases. Each simulation case is replicated $500$ times to estimate the mean received signal at the receiver side.

What is of prime interest for modeling an MCvD channel is to model, as noted above, the received signal (i.e., the number of hitting molecules until time $t$). To model the received signal, we utilize curve-fitting and artificial neural network techniques. In the performance figures, we cannot present all 135 cases, but we offer some example scenarios and average RMSE plots.

\subsection{Received Signal Analysis}
\begin{figure}[!t]
\begin{center}
	\includegraphics[width=1.0\columnwidth,keepaspectratio]%
	{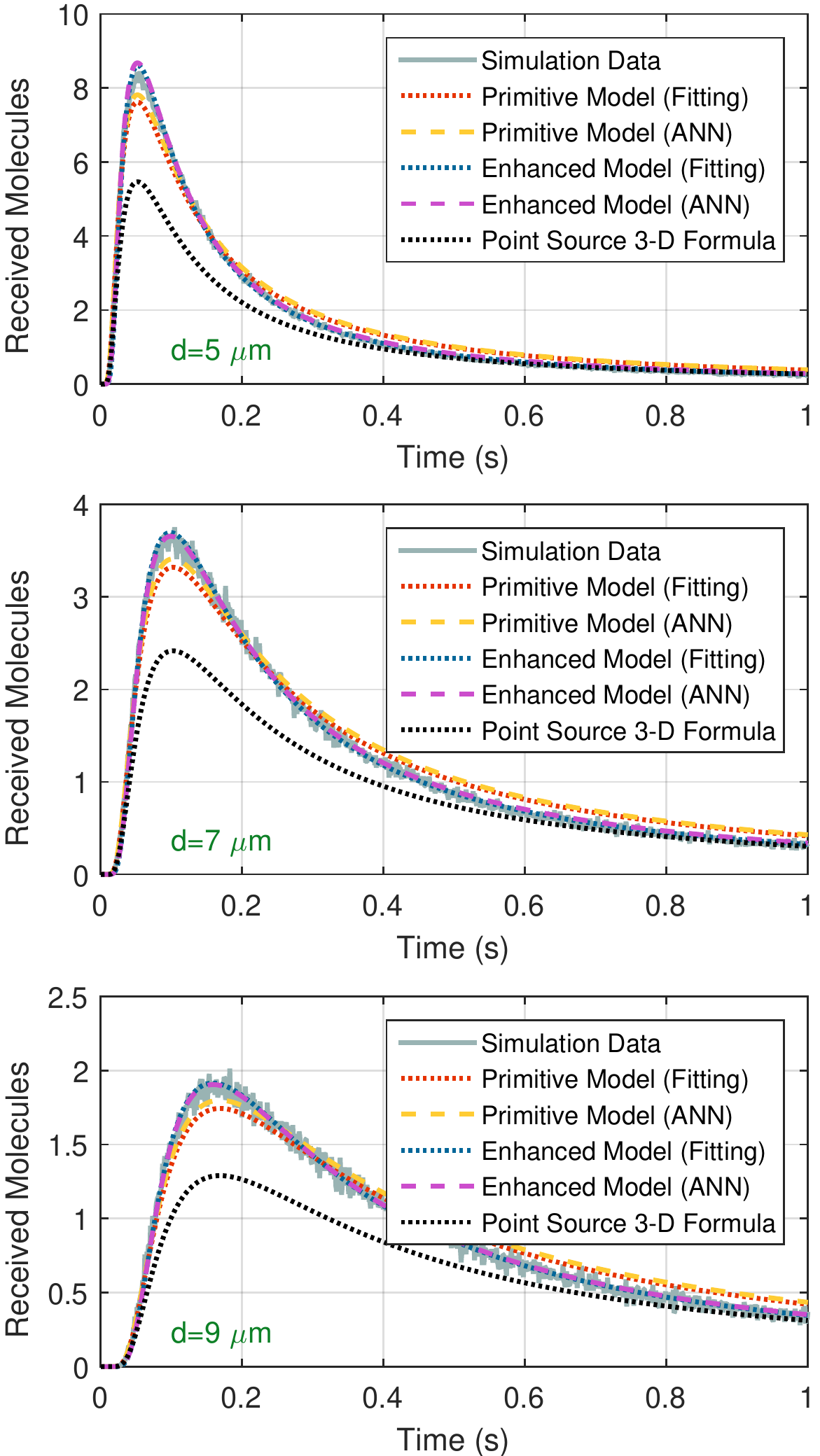}
	\caption{Received signal plots for $\rtx\!=\!\SI{4}{\micro\meter}$, $\rrx\!=\!\SI{8}{\micro\meter}$, $D\!=\!\SI{80}{\micro\meter^2/\second}$ with time resolution of $\SI{1}{\milli\second}$. From top to bottom, plots correspond to $d\!=\!\SI{5}{\micro\meter}$, $d\!=\!\SI{7}{\micro\meter}$ and $d\!=\!\SI{9}{\micro\meter}$ cases.}
	\label{fig____received_signal}
\end{center}
\end{figure}

In Fig.~\ref{fig____received_signal}, the received signal is plotted for simulation data, curve-fitting, and ANN techniques. Note that the ANN technique requires no simulation data, while the curve-fitting method does. After training an ANN, we estimated channel model parameters for the validation data by giving only the system parameters as input. For the \emph{primitive model} that is given in~\eqref{eqn_model_primitive}, a single channel model parameter was estimated. On the other hand, for the \emph{enhanced model} that is given in~\eqref{eqn_model_enhanced}, three channel model parameters were estimated. 

It can be clearly seen that the \emph{enhanced model} fits the simulation data better than does the \emph{primitive model}. At the peak and the tail part of the received signal, we observe that the  \emph{enhanced model} outperforms the \emph{primitive model} and the point transmitter formulation given in~\eqref{eqn_3d_frac_received_point_src}. The second observation suggests that, with increased distance, the estimation performance of the received signal is improved. This observation is also supported with the RMSE plots in Fig.~\ref{fig____rmse_plots}. Another observation is that the trained ANNs generalize the fitted model parameters well. In other words, we observe that the ANN curves are close to the curves that are produced by curve fitting. Without knowing the simulation data (by using only the system parameters as inputs), the trained ANN estimates the channel model parameters well.

\subsection{SIR Analysis}
For a more in-depth analysis we used the following metric, SIR, formulated as:
\begin{align}
\begin{split}
SIR(t) = \frac{F_\text{hit}^{3\text{D}} (t)}{F_\text{hit}^{3\text{D}} (\tend)-F_\text{hit}^{3\text{D}} (t)} 
\end{split}
\label{eqn_sir_defn}
\end{align}
where we can also substitute $F_\text{hit}^{3\text{D}} (t)$ with the channel model functions given in~\eqref{eqn_model_primitive} and~\eqref{eqn_model_enhanced}. This metric basically represents the ratio of the cumulative number of received molecules until time $t$ and the number of interference molecules. SIR plots are important since they show the performance of modeling the ratio of desired signal and interference.
\begin{figure}[!t]
\begin{center}
	\includegraphics[width=0.9\columnwidth,keepaspectratio]%
	{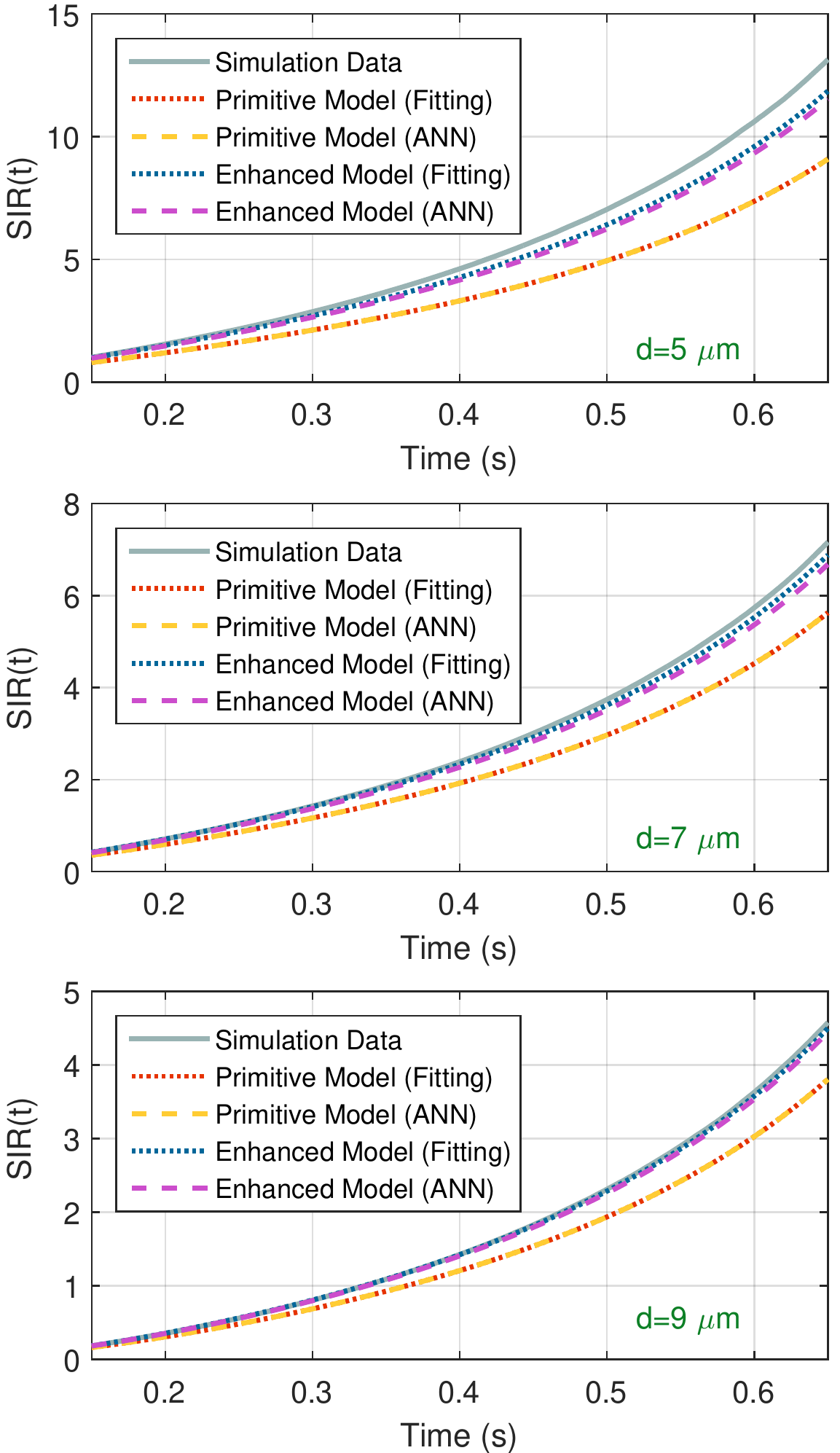}
	\caption{SIR plots for $\rtx\!=\!\SI{4}{\micro\meter}$, $\rrx\!=\!\SI{8}{\micro\meter}$, $D\!=\!\SI{80}{\micro\meter^2/\second}$ with time resolution of $\SI{1}{\milli\second}$. From top to bottom, plots correspond to $d\!=\!\SI{5}{\micro\meter}$, $d\!=\!\SI{7}{\micro\meter}$ and $d\!=\!\SI{9}{\micro\meter}$ cases.}
	\label{fig____sir_plots}
\end{center}
\end{figure}
In Fig.~\ref{fig____sir_plots}, SIR is plotted for the simulation data, curve-fitting, and ANN techniques. The \emph{enhanced model} estimates the received signal better than the \emph{primitive model}. Moreover, the longer the distance the better the channel parameter estimation performance. For $d\!=\!\SI{9}{\micro\meter}$ case, the \emph{enhanced model} is very close to the simulation data. However, the \emph{primitive model} with one scaling factor does not adequately model the simulation data. Again, we see that the trained ANN generalizes the fitting method well for both of the cases (i.e., fitting and ANN curves are overlapping for both models). After $d\!=\!\SI{7}{\micro\meter}$, ANN performs close to the simulation data.

\subsection{RMSE Analysis}
To better understand the performance of the estimation technique, we evaluated the mean RMSE of cases with respect to (wrt) simulation data in terms of the number of received molecules until time $t$. For RMSE analysis, we grouped results with respect to distance and $\rrx$ so that each group was the average of nine cases.
\begin{figure}[!t]
\begin{center}
	\includegraphics[width=1.0\columnwidth,keepaspectratio]%
	{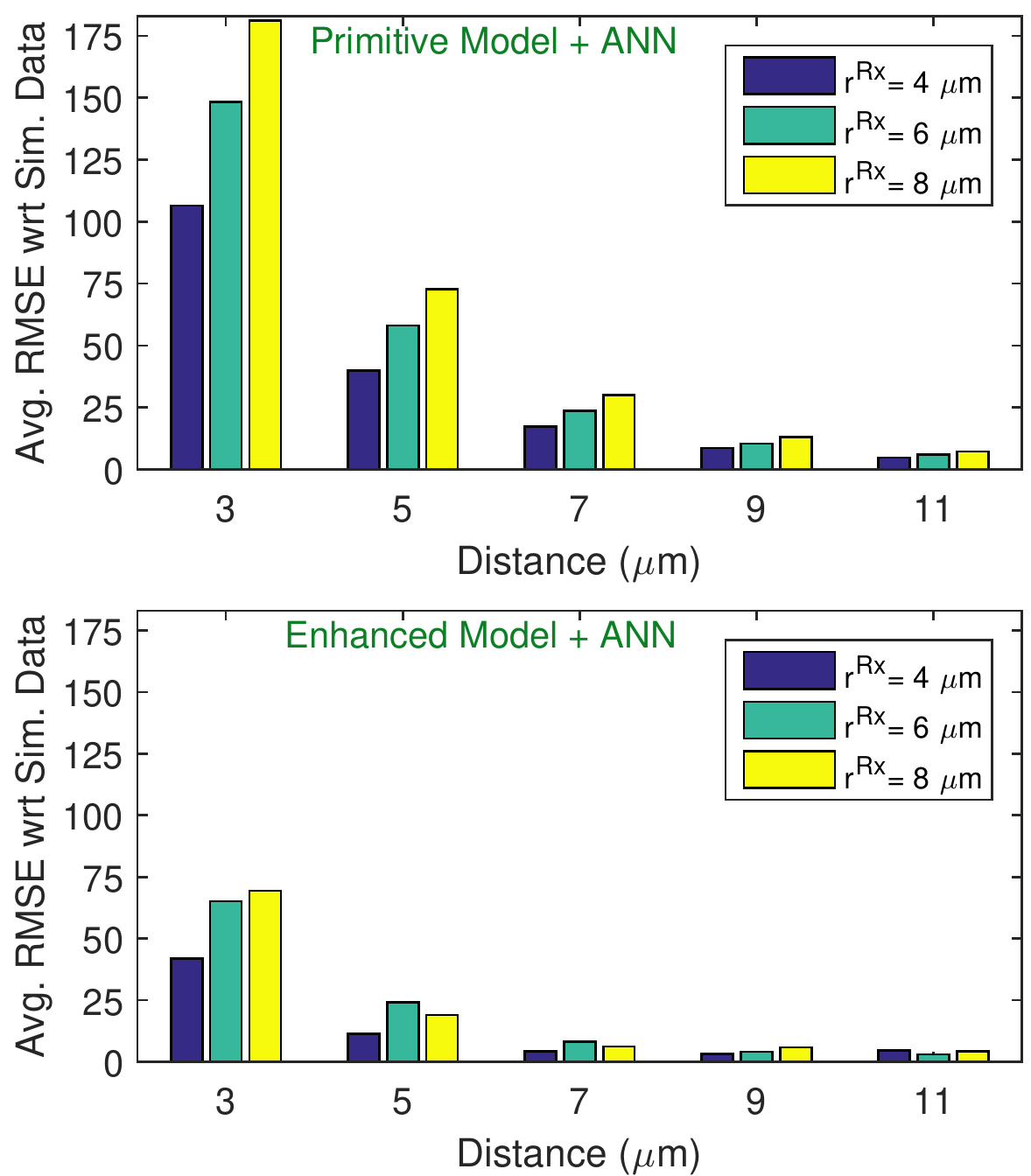}
	\caption{Average RMSE plots of ANN estimation with primitive and enhanced models.}
	\label{fig____rmse_plots}
\end{center}
\end{figure}

In Fig.~\ref{fig____rmse_plots}, the RMSEs of the channel parameter estimation methods are presented for ANNs with \emph{primitive} and \emph{enhanced} models. The first observation suggests that the estimation performance gets better with increasing distance. Moreover, the RMSE of the \emph{enhanced model} is significantly lower than that of the \emph{primitive model}. Note that the ANN technique requires no simulation data, utilizing only the system parameters as input to estimate the channel parameters. The \emph{primitive model} scales the formulation of point transmitter case and the larger $\rrx$ case deviates more from the point transmitter case, as the probability rises of receiving obstructed and reflected molecules. Therefore, for the \emph{primitive model}, performance of the estimation of the received signal is better in terms of RMSE for smaller $\rrx$. 

\section{Conclusion}
In this work, we developed a novel technique to model the received signal in MCvD with a spherical transmitter. In the literature, a point transmitter is assumed for the tractability of the mathematical derivations in a first-hitting process framework. Approaching the problem from a unique perspective, we utilized an artificial neural network technique and a model function for the number of received molecules. After training an ANN, we were able to ask the ANN to estimate the channel model parameters for different system setups. Our proposed technique has promising results for modeling the number of received molecules until time $t$. We observed that the proposed technique models the received signal and SIR more effectively for longer distances. The proposed technique may be utilized to model an MCvD channel in other studies that assumes a spherical transmitter instead of a point transmitter.

\section*{Acknowledgment}
This research has been supported by the MSIP (Ministry of Science, ICT and Future Planning), Korea, under the ``ICT Consilience Creative Program" (IITP-R0346-16-1008) supervised by the IITP (Institute for Information \& Communications Technology Promotion) and by the Basic Science Research Program (2014R1A1A1002186) funded by the MSIP, Korea, through the National Research Foundation of Korea.


\bibliographystyle{IEEEtran}
\bibliography{mybibs_sph_src}

\end{document}